\documentstyle[aps,preprint,tighten]{revtex}

\begin{document}

\title{Charge Density Wave Ratchet}

\author{Mark I. Visscher and Gerrit E.W Bauer}
\address{Theoretical Physics Group, Department of Applied Physics and DIMES \\
Delft University of Technology, Lorentzweg 1, 2628 CJ Delft, The Netherlands}

\date{Received date: March 29, 1999}

\maketitle

\begin{abstract}
We propose to operate a locally-gated charge density wave as an
electron pump. Applying an oscillating gate potential with
frequency $f$ causes equally spaced plateaux in the sliding charge density wave
current separated by $\Delta I=2eNf,$ where $N$ is the number 
of parallel chains. The effects of thermal noise are investigated.
\end{abstract}


A metallic gate electrode on the surface of charge density wave (CDW) compounds can cause transistor
action by modulating the threshold field for depinning  of the collective CDW mode \cite{adelman}.
Several suggested microscopic mechanisms fail to explain the observed large asymmetric gate modulation. 
Nevertheless, we may expect that the sliding mode can be manipulated even more
effectively on thin-films of charge density wave materials which have been grown recently \cite{zant}. 
Here we propose an electron-pump based on structured CDW films, which we call a ``CDW
ratchet".

In the early nineties the ``single-electron turnstile'' was realized in both
metals \cite{geerligs} and semiconductors \cite{kouwenhoven}. In these 
structures electrons are transferred one by one by using time-dependent gate voltages
to modulate the Coulomb blockade or by alternatively raise and lower tunnel
barrier heights. The dc current through such a device scales linearly with
the applied frequency $I=ef$, and the current-(bias)voltage characteristics
shows equally spaced current plateaux. The single
electron pump may find an application as a current standard, since the
oscillation frequency $f$ can be controlled to high precision. The accuracy of the 
current is affected by numerous error mechanisms as offset charges, cycle missing, 
thermal fluctuations, or co-tunneling 
\cite{pothier,jensen}. Some of these errors may be suppressed 
in linear arrays of tunnel junctions \cite{bakhvalov} or in periodically gated quantum wires \cite{niu}, but 
devices with a charge density wave ground state should not  
suffer from these drawbacks at all.

Charge density waves occur in quasi one-dimensional metals, like {\rm NbSe}$%
_{3}$ or {\rm K}$_{0.3}${\rm MnO}$_{3}$ \cite{gorkov}. Below a critical temperature $T_{c}$ 
($183 K$ for $\rm{K_{0.3}MoO_{3}}$), the ground state consists of a lattice distortion coupled to an electron
density modulation $n_{CDW}\sim |\Delta (x,t)|\cos \left[ 2k_{F}x+\chi
(x,t)\right] ,$ where $2|\Delta |$ is the gap in the quasi-particle spectrum
and the phase $\chi $ denotes the position of the CDW relative to the crystal lattice.
Incommensurate CDW's support a unique sliding mode of transport above a
weak threshold field. This collective motion of the CDW carries electrical
current proportional to $\partial _{t}\chi $ and is the source of
narrow-band noise and non-linear conductance characteristics. The threshold
field arises from the interaction of the CDW with defects or impurities in
the system, which can, as mentioned above, be manipulated by external gate
electrodes \cite{adelman}.

As experimental setup for the CDW electron pump we envisage a thin strip
of CDW material consisting of $N$ chains with dimensions of the order of the
Fukuyama-Lee-Rice coherence lengths ($\xi_{\parallel}$ is typically micrometers and 
$\xi_{\parallel}/\xi_{\perp}\sim 10-100$ \cite{fukuyama}), such that the CDW is characterized by a
single degree of freedom. A thin metallic gate electrode separated by an
insulating layer is placed on top, perpendicular to the CDW chains, and connected to
an oscillating voltage. Alternatively, one could also think of the tip of an
STM as gate electrode. The dynamics of the CDW with an oscillating time-dependent 
gate potential can be described by the classical equation of motion for the phase 
$\chi (t)$ in the single particle model \cite{gruner}

\begin{equation}
\eta \frac{\partial \chi }{\partial t}+V_{p}(\chi ;t)=V_{b}+V_{n}(t)
\label{motion}
\end{equation}
where $\eta =\hbar R_{c}/eR_{Q}$, $R_{c}$ is a damping resistance and $%
R_{Q}=h/2Ne^{2}$ is the $N$-mode quantum resistance. The pinning potential $%
eV_{p}$ is periodic in the phase and contains an explicit time dependence $%
V_{p}(\chi +2\pi ;t+2\pi /\omega )=V_{p}(\chi ;t)$, where $\omega =2\pi f$ 
is the frequency of the oscillating gate potential. The driving term consists
of the bias voltage $V_{b}$ and a thermal (Nyquist) noise term $V_{n}(t)$ with $<V_{n}(t)>=0$ and 
$<V_{n}(t)V_{n}(t^{\prime })>=2\eta k_{B}T\delta (t-t^{\prime })/e$. We will disregard the inertia since CDW's are in general strongly overdamped due to
their large effective mass. Note that there is no oscillating drive term,
which is known to lead to phase-locking and Shapiro steps in the
current--voltage characteristic. The interpretation of Eq. (\ref{motion}) is
straightforward: a thermally activated classical particle moving in a tilted
washboard potential, of which the amplitude changes periodically in time. The
applied electric field directs the motion of the CDW, and the oscillating
gate guides the CDW downwards, thus causing
an electric current. 

In the presence of thermal noise, the dynamics of the CDW can be described
by a Fokker-Planck equation for the probability density $P(\chi ,\tau )$ of
finding the phase $\chi $ in the interval $\chi +d\chi $ at time $\tau $ 
\cite{kramers}$.$ In the following we will assume that the pinning potential
may be approximated by its lowest harmonics: 
\begin{equation}
V_{p}(\chi ;t)=(V_{T}+\alpha V_{g}\sin \omega t)\sin \chi ,
\label{pinpot}
\end{equation}
where $V_{T}$ is a constant threshold potential, which is 
modulated by the fraction $\alpha $ of the oscillating gate potential $V_{g}.
$ In this case the Fokker-Planck equation reads 
\begin{equation}
\frac{\partial P}{\partial \tau }=D\frac{\partial ^{2}P}{\partial \chi ^{2}}%
+(1+\tilde{V}_{g}\sin \tilde{\omega}\tau )\frac{\partial }{\partial \chi }%
(\sin \chi P)-\tilde{V}_{b}\frac{\partial P}{\partial \chi },  \label{fokker}
\end{equation}
where we introduced the dimensionless parameters $\tilde{V}_{g}=\alpha
V_{g}/V_{T}$, $\tilde{V}_{b}=V_{b}/V_{T}$, $D=k_{B}T/eV_{T}$, and $\tilde{%
\omega}=\eta \omega /V_{T}$. The initial condition at time $\tau =\tau _{0}$
is given by $P(\chi ,\tau _{0})=\delta (\chi -\chi _{0}).$ The problem is
now reduced to the diffusion of a classical particle in a time dependent periodic
potential. By substituting the Fourier series 
\begin{equation}
P(\chi ,\tau )=\sum_{n=-\infty }^{\infty }P_{n}(\tau )e^{-in\chi }
\end{equation}
into Eq. (\ref{fokker}) we obtain the equation for the Fourier components $%
P_{n}(\tau )$%
\begin{eqnarray}
& &\frac{\partial P_{n}}{\partial \tau }= \nonumber \\ 
& & (-Dn^{2}+in\tilde{V}_{b})P_{n}-(1+%
\tilde{V}_{g}\sin \tilde{\omega}\tau )\frac{n}{2}[P_{n+1}-P_{n-1}].
\label{fourier}
\end{eqnarray}
The total dc current $I$ through the system is defined as 
\begin{equation}
I=\frac{V_{b}}{R_{c}}-\frac{1}{2iR_{c}}<(V_{T}+V_{g}\sin \tilde{\omega}\tau
)[P_{1}(\tau )-P_{-1}(\tau )]>,  \label{current}
\end{equation}
where the brackets denote time averaging.

We first consider the $T=0$ (noiseless) limit (Eq. (\ref{motion})). Without gate voltage the current is obviously given by $I=0$
for $V_{b}<V_{T}$ and $I=\sqrt{V_{b}^{2}-V_{T}^{2}}/R_{c}$ for $V_{b}\geq
V_{T}.$ Figure \ref{fig:IV} shows the $I-V_{b}$ characteristic for different values of the
external frequency $f$. Distinct current plateaux appear in the $I-V_{b}$
below the normal threshold potential. Each plateau corresponds to 
the displacement of a quantized number of wave fronts in one cycle.
The steps are equally separated by $%
\Delta I=2eNf$, where the factor $2$ reflects spin-degeneracy.
Far above threshold the differential
resistance approaches its normal value $R_{c}$. Furthermore, the
current-frequency relation has a fan structure, as is shown in Fig. 
\ref{fig:fan}. The current scales linearly with frequency 
\begin{equation}
I=2emNf,
\end{equation}
where the integer $m$ is the number of displaced wave lengths in one cycle. 
In the limit $\omega \rightarrow 0$ the current approaches the constant 
value as
\begin{equation}
I(\omega\rightarrow 0)=\frac{1}{2\pi R_{c}}{\rm{Re}}\int_{0}^{2\pi} d\tau \sqrt{V_{b}^{2}-(V_{T}+
V_{g}\sin\tau)^{2}}.
\end{equation}
The current drops to zero at the
frequency which corresponds to the time to displace 
the CDW by one wave length. 

Next we investigate the effects of the thermal noise at finite temperatures.
In the case where the external oscillating gate
potential is absent, the stationary state solution to Eq. (%
\ref{fourier}) is easily calculated as 
\begin{equation}
P_{n}(\tau \rightarrow \infty )=\frac{I_{n-iz_{0}}(z)}{I_{-iz_{0}}(z)},
\label{stationary}
\end{equation}
where $I_{q}(z)$ is a modified Bessel function with imaginary argument $q$, $%
z_{0}=\tilde{V}_{b}/D$ and $z=1/D$. Using the relation $P_{n}^{*}=P_{-n}$,
the CDW current is obtained from Eq. (\ref{current}) 
\begin{equation}
I=\frac{V_{b}}{R_{c}}-\frac{V_{T}}{R_{c}}{\rm{Im}}\frac{I_{1-iz_{0}}(z)}{%
I_{-iz_{0}}(z)}.
\end{equation}
This static result is exactly analogous to the case of strongly overdamped
Josephson junctions with an external noise current \cite
{ambegaokar,ivanchenko}. The main effect of finite temperatures is a
smoothening of the square-root threshold singularity near $V_{T}$ and an
exponentially small (for $1/D\gg 1$) but nonzero conductance as $%
V_{b}\rightarrow 0$. In the presence of the oscillating gate we solve Eq. (%
\ref{fourier}) numerically. In Fig. \ref{fig:temp} the $I-V_{b}$ curves at
external frequency $\tilde{\omega}=0$ and $\tilde{\omega}=0.4$ are shown for
different temperatures $D=k_{B}T/eV_{T}.$ As expected, finite temperatures
smear out the sharp transitions between the current plateaux. This is more
pronounced at larger bias, since the escape rate due to the thermal
fluctuations increases. At even higher temperatures $1/D \lesssim 5$ the plateaux disappear
and the resistance becomes linear $V_{b}=IR_{c}$.

For operation as a current standard, an individual CDW ratchet must first be gauged to 
determine the number of parallel chains $N$. Once known, the linear dependence of the
current on $N$ in principle improves the accuracy of the current quantization as compared 
to single electron pumps. The band-width of
single-electron turnstile devices is limited by the competition between
large detection currents (large frequencies) and low noise levels (low
frequencies). The present CDW device, however, allows for large
currents at low frequencies, and is 
robust to the error sources of the single-electron pump.

Since CDW's have a large single-particle energy gap $2|\Delta|$, the quasi-particle
contribution to the current is of the order $\exp (-2|\Delta| /k_{B}T)$ and
can be neglected for sufficiently low temperatures $T \ll T_{c}$. The gap also makes the
system robust to static disorder and single-particle quantum fluctuations in the total charge per unit
cell. Phase-slip processes at the current contacts can be avoided in a four-terminal
measurement, where the voltage probes are located far from the current
source and drain. Higher harmonics of the pinning potential Eq. (\ref{pinpot}) will give  
corrections only near the thresholds. The elasticity of the CDW
can phenomenologically be described by geometrical capacitances $C$
of the leads, which define the typical Coulomb energy for phase
deformations. The assumption of a rigid CDW
is then justified in the limit of weak pinning and low temperatures, such that $|\Delta| 
\gg e^{2}/C \gg eV_{T} \gg k_{B}T $ \cite{visscher}. 
We disregarded the role of macroscopic quantum tunneling
of the CDW, which we consider less important than coherent co-tunneling in
single-electron devices because of the high effective mass. 

We conclude by summarizing our results. We present the idea of a locally
gated CDW as an electronic ratchet. An oscillating gate potential causes
equidistant current plateaux. The current
scales linearly with the external frequency and the number of chains. The
coherent electronic ground state of CDW's and the intrinsic property of
parallel conducting chains makes this device a serious candidate for the
current standard with possible higher accuracy than single electron devices,
even in the presence of thermal noise.

It is a pleasure to thank Behzad Rejaei, Michel Devoret, Erik Visscher, Herre van der Zant,
and Yuli Nazarov for stimulating discussions. This work is part of the research program for the ``Stichting voor
Fundamenteel Onderzoek der Materie'' (FOM), which is financially
supported by the ''Nederlandse Organisatie voor Wetenschappelijk
Onderzoek'' (NWO).  This study was supported by the NEDO joint
research program (NTDP-98).

\begin{figure}[t]
\caption{Current-voltage
characteristic of the CDW ratchet at $T=0$ for external frequencies $\tilde{%
\omega}=0;0.1;0.5$. The current plateaux are equally spaced by $\Delta
I=2eNf.$}
\label{fig:IV}
\end{figure}

\begin{figure}[t]
\caption{Current-frequency fan at bias voltages
$\tilde{V}_{b}=0.4$ and $0.5$.}
\label{fig:fan}
\end{figure}

\begin{figure}
\caption{Current-voltage characteristic of the CDW ratchet
including thermal noise for different temperatures, $eV_{T}/k_{B}T=\infty
,100,20,10$.}
\label{fig:temp}
\end{figure}


\begin{thebibliography}{99}
\bibitem{adelman}  T.L. Adelman, S.V. Zaitsev-Zotov, and R.E. Thorne, Phys.
Rev. Lett. {\bf 74}, 5264 (1995).

\bibitem{zant}  H.S.J. van der Zant, O.C. Mantel, C. Dekker, J.E. Mooij, and
C. Treaholt, Appl. Phys. Lett. {\bf 68}, 3823 (1996).

\bibitem{geerligs}  L.J. Geerligs, V.F. Anderegg, P.A.M. Holweg, J.E. Mooij,
H. Pothier, D. Est\`{e}ve, C. Urbina, and M.H. Devoret, Phys. Rev. Lett. 
{\bf 64}, 2691 (1990).

\bibitem{kouwenhoven}  L.P. Kouwenhoven, A.T. Johnson, N.C. van der Vaart,
and C.J.P.M. Harmans, Phys. Rev. Lett. {\bf 67}, 1626 (1991).

\bibitem{gruner}  G. Gr\" uner, A. Zawadowski, and P.M. Chaikin, Phys. Rev.
Lett. {\bf 46}, 511 (1981).

\bibitem{pothier}  H. Pothier, P. Lafarge, D. Esteve, C. Urbina, and M.H.
Devoret, IEEE Trans. Instr. Meas. {\bf 42}, 324 (1993).

\bibitem{jensen}  H.D. Jensen, and J.M. Martinis, Phys. Rev. B {\bf 46,}
13407 (1992).

\bibitem{bakhvalov} N.S. Bakhvalov, G.S. Kazacha, K.K. Likharev, and S.I. Serdyukova,
Sov. Phys. JETP {\bf 68}, 581 (1989).  

\bibitem{niu}  Q. Niu, Phys. Rev. Lett, {\bf 64}, 1812 (1990).

\bibitem{gorkov} For a review see {\it Charge Density Waves in Solids},
edited by L.P. Gor'kov and G. Gr\" uner (North-Holland, Amsterdam, 1989).

\bibitem{fukuyama} H. Fukuyama and P.A Lee, Phys. Rev. B {\bf 17}, 535 (1977).
P.A. Lee and T.M. Rice Phys. Rev. B {\bf 19},3970 (1979).

\bibitem{kramers}  H.A. Kramers, Physica {\bf 7}, 284 (1940); Shang-keng
Ma in {\it Modern theory of Critical Phenomena} (Addison-Wesley, Redwood City (Ca),
1976).

\bibitem{ambegaokar}  V. Ambegaokar and B.I. Halperin, Phys. Rev. Lett. {\bf %
22}, 1364 (1969).

\bibitem{ivanchenko}  Yu. M. Ivanchenko and L.A. Zil'berman, Sov. Phys. JETP 
{\bf 28}, 1272 (1969).

\bibitem{visscher} M.I. Visscher, {\it Transport in Mesoscopic Charge Density Devices} (Delft University
Press, Delft, 1998).

\end{thebibliography}
\end{document}